\documentclass[reprint,superscriptaddress,aps,prb,twocolumn]{revtex4-1}
\usepackage[utf8]{inputenc}
\usepackage{textcomp} 
\usepackage{setspace}
\usepackage{amsmath}
\usepackage{breqn}
\usepackage{graphicx}
\usepackage{verbatim}
\usepackage{amsfonts}
\usepackage{amssymb}
\usepackage{xcolor}
\usepackage{physics}
\usepackage{siunitx}
\usepackage{gensymb}

\usepackage[colorlinks,linkcolor=blue,anchorcolor=blue,citecolor=blue,urlcolor=black]{hyperref}
\DeclareGraphicsExtensions{.pdf,.eps,.png,.jpg,.mps}

\begin{document}

\title{On-chip optical twisted bilayer photonic crystal}
\author{Haoning Tang$^{1\bot \ast}$, Beicheng Lou$^{2\bot }$, Fan Du$^{1}$,  Mingjie Zhang$^{1}$, Xueqi Ni$^{1}$, Weijie Xu$^{1}$, Rebekah Jin$^{3}$, Shanhui Fan$^{2\ast}$, Eric Mazur$^{1\ast}$\\
\vspace{3pt}
$^1$School of Engineering and Applied Sciences, Harvard University, Cambridge, MA 02138, USA\\
$^2$Department of Applied Physics and Ginzton Laboratory, Stanford University, Stanford, CA 94305, USA\\
$^3$University of California, Los Angeles, CA 90095, USA\\
\vspace{3pt}
$^\bot$Authors contributed equally to this work.\\
$^\ast$Corresponding author: hat431@g.harvard.edu, shanhui@stanford.edu, mazur@seas.harvard.edu}

\maketitle

\noindent{\bf Recently, moir\'e engineering has been extensively employed for creating and studying novel electronic materials in two dimensions. However, its application in nanophotonic systems has not been widely explored so far. Here, we demonstrate that twisted bilayer photonic crystals provide a new photonic platform with twist-angle-tunable optical dispersion. Compared to twisted two-dimensional materials, twisted bilayer photonic crystals host a rich set of physics and provide a much larger number of degrees of freedom — choice of material, lattice symmetry, feature size, twist angle, and interlayer gap, — which promises an unprecedented toolbox for tailoring optical properties.  We directly visualize the dispersion throughout the optical frequency range and show that the measured optical response is in good quantitative agreement with numerical and analytical results. Our results reveal a highly tunable band structure of twisted bilayer photonic crystals due to moir\'e scattering. This work opens the door to exploring unconventional physics and novel applications in photonics.
}

There are emerging interests in using moiré physics to engineer optical dispersion. For example, moiré-patterned single-layer\cite{Wang2020-ce, Fu2020-zh, Talukdar2022-jf, Mao2021-ye, Zhou2020-ox, Shang2021-kh, Lin2022-cz, Han2015-ak, Asboth2016-cd, Zeng2021-po, Alnasser2021-nq, Kartashov2021-kl, Khurgin2000-au, Beravat2016-vd, Wong2012-kx} and twisted-bilayer\cite{Lou2022-fp, Lou2021-dv, https://doi.org/10.48550/arxiv.2211.07230, Wang2022-nf, Dong2021-mh, Lubin2013-fu, Wu2018-kb, Yi2022-nh, Nguyen2022-bt, Huang2022-yp, Hu2021-ty, Sunku2018-uf, Duan2020-yw, Chen2020-lf, Krasnok2022-bf, Zhang2021-qg, Hu2021-mk, Chen2021-ef, Wu2018-on, Liu2022-ru} photonic structures exhibit ultra-flat bands with no dispersion. The moiré pattern created by twisting two photonic structures relative to each other gives rise to distinctive optical properties, including nonlinear enhancement\cite{Yao2021-ma} and anisotropic dispersion\cite{Hu2020-lh}. Using a pair of photonic crystal slabs\cite{Joannopoulos2008-eq} that are twisted relative to each other provides a large number of degrees of freedom — choice of material, lattice symmetry, feature size, twist angle, and interlayer gap, — and permits tailoring the optical properties of the material. In particular, recent theoretical work shows that twisted bilayer photonic crystal (TBPhC) structures exhibit slow light\cite{Tang2021-bx}, facilitating the study of strong light-matter interactions and Purcell enhancement\cite{Tang2022-yh}, and frequency filtering\cite{Lou2022-pd}. To date, however, there has not been any demonstration of TBPhC devices in the optical frequency range. 

In this paper, we report on the fabrication of dielectric TBPhC structures that work in the optical frequency range and on the measurement of their momentum-space-resolved optical response and optical band structure. We also compare the band-folding and band-hybridization phenomena observed in the measurement to numerical and analytical results. The results presented in this paper open the door to experimentally exploring theoretically-predicted unconventional physics, such as nontrivial topological phenomena\cite{https://doi.org/10.48550/arxiv.2211.07230} and bound-states-in-the-continuum\cite{PhysRevLett.128.253901}. They also allow potential applications such as adaptive\cite{Gan2012-ne, Wang2014-af} and compressive\cite{Gan2012-ne, Wang2014-af} sensing, and on-chip, real-time configurable optical filters, polarimeters, switches, and lasers.

\begin{figure*}[ht]
\centering
\includegraphics[width=17cm]{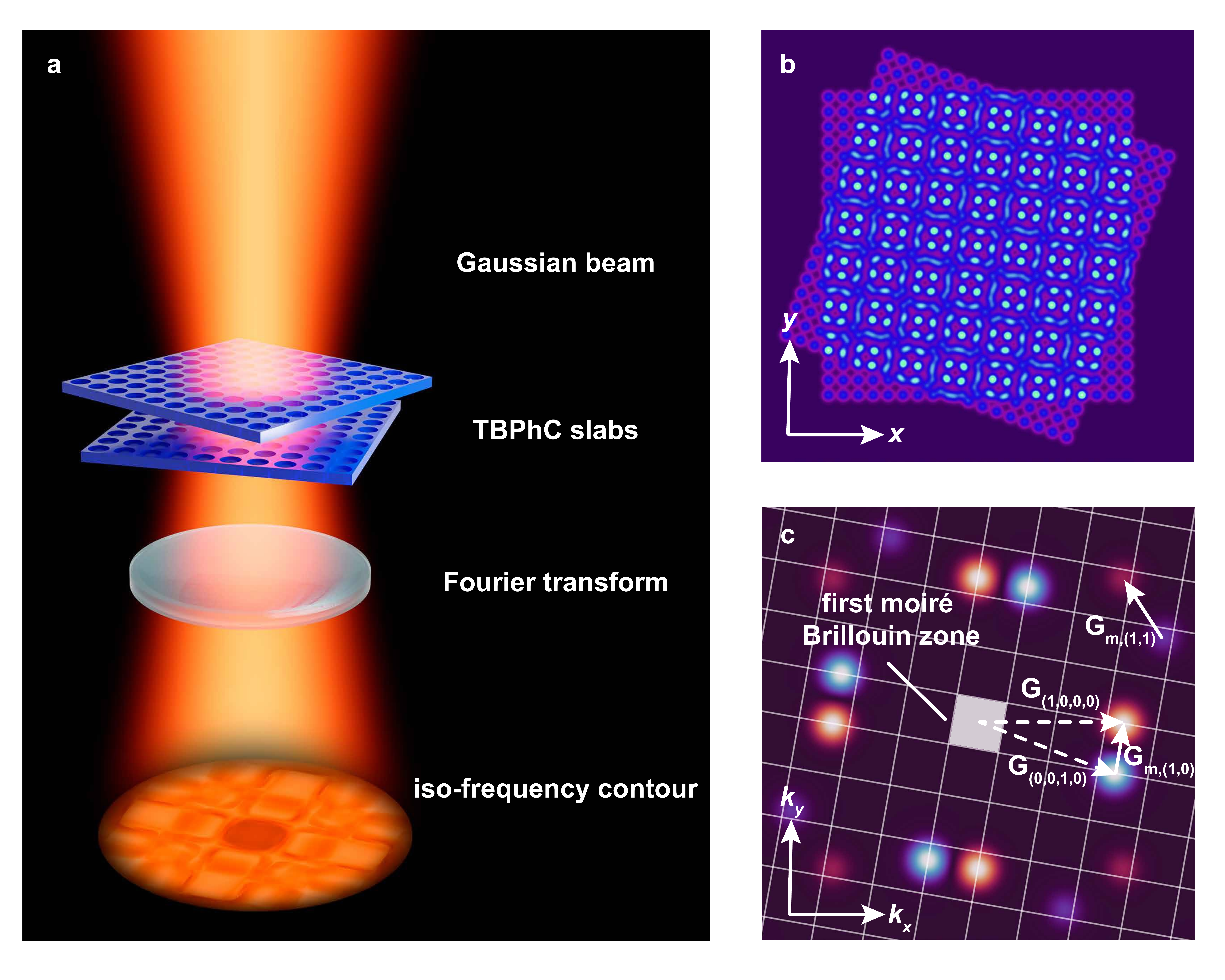}
\caption{\textbf{Schematic of the experiment and the process of moiré scattering. a}, Twisted bilayer photonic crystal slabs with circular holes in a square lattice illuminated by focusing Gaussian incident light and mapping the transmitted light to its momentum space. \textbf{b}, Moiré lattice from top-down view. \textbf{c}, The twisted bilayer reciprocal lattice and basis are involved in the scattering process between the two layers. The reciprocal spaces of the first and second layer are marked by the center of the blue and orange spots, respectively. Larger and brighter spots indicate stronger scattering. The square mesh and shaded area indicate the first-order moiré Brillouin zone. A theoretical analysis of the moiré scattering strength is in Supplementary Material Section 2.
}
\label{fig1}
\end{figure*}

\begin{figure*}[ht]
\centering
\includegraphics[width=17cm]{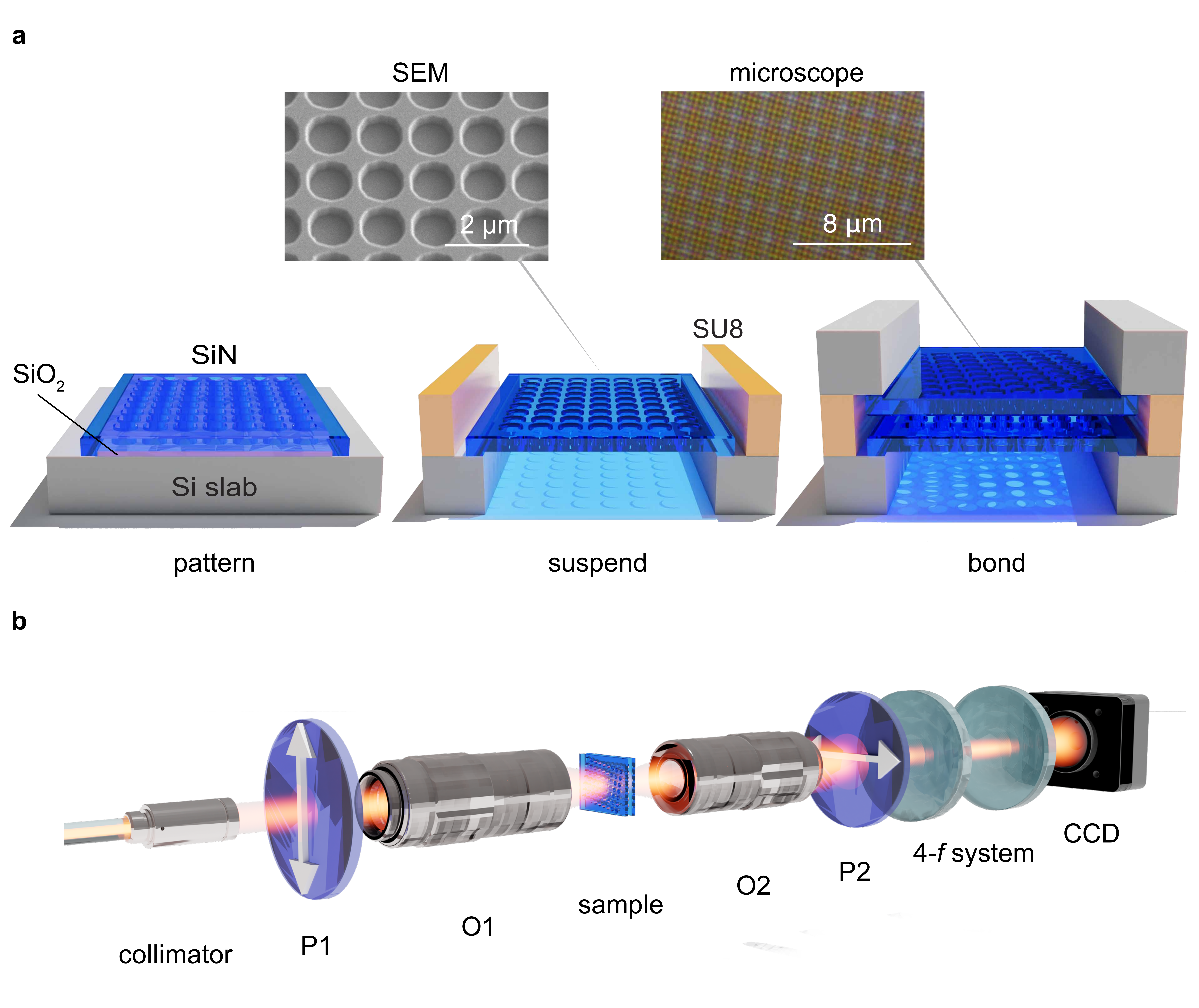}
\caption{\textbf{Device fabrication process and experiment setup. a},  The fabrication process of the twisted bilayer photonic crystal slabs. The zoomed in pictures are a scanning electron microscope (SEM) image of fabricated single-layer PhC slabs and a 50x microscope image of TBPhC slabs with a visible moiré pattern. \textbf{b}, A schematic of the measurement setup. The red light line represents the incident light and its direct transmission and radiation-induced scattering from the bilayer lattice. P, polarizer.  O, objective lens. CCD, charge-coupled device. 
}
\label{fig2}
\end{figure*}

\begin{figure*}[ht]
\centering
\includegraphics[width=17cm]{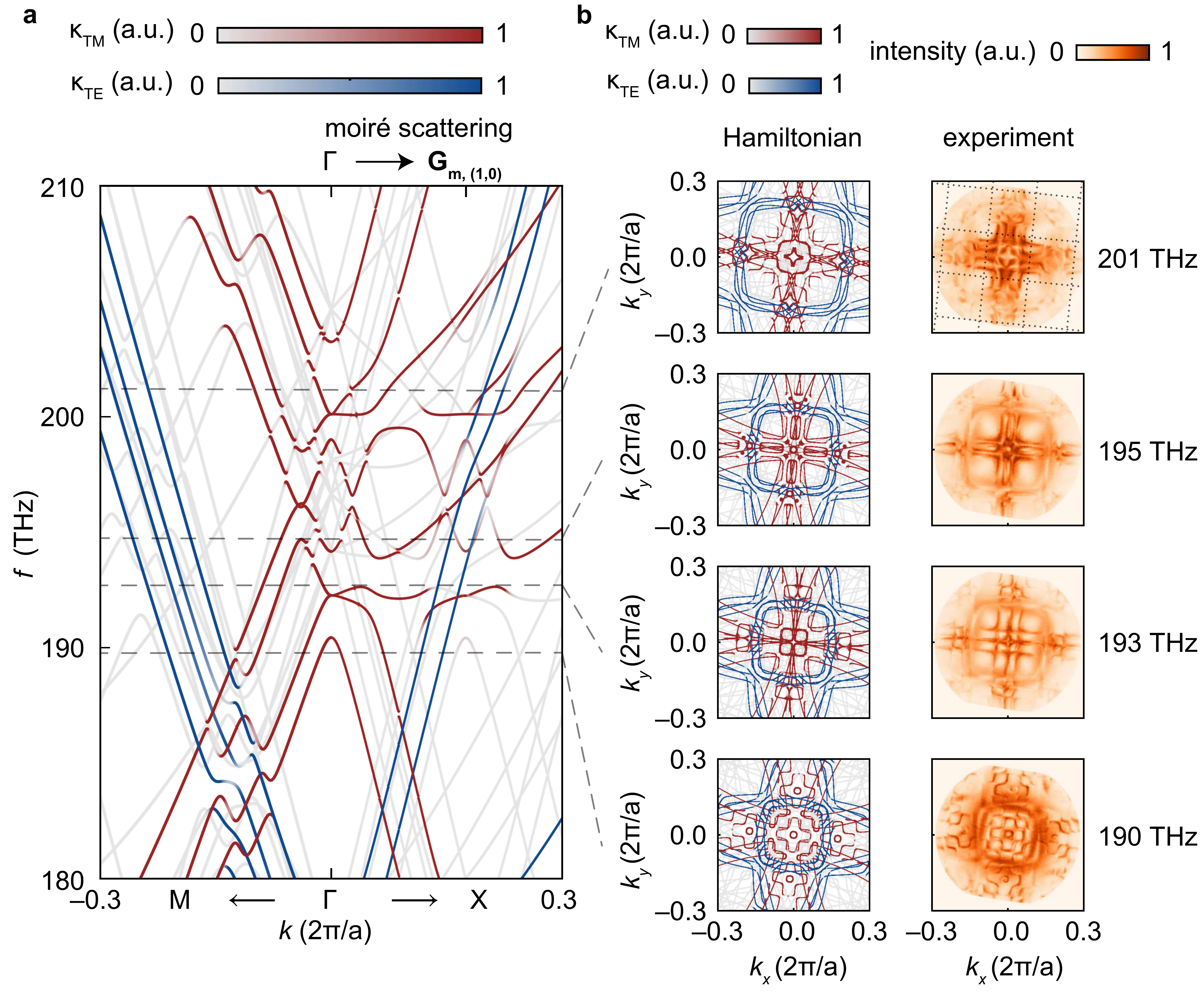}
\caption{\textbf{Hamiltonian band structure compared with the iso-frequency contours in measurement. a}, the Hamiltonian band structure of the TBPhC slabs with a twist angle of  $\theta=10.0\degree$. TM-like and TE-like modes are shown in red and blue, with each color bar indicating the coupling strength. The dashed lines indicate the frequencies of iso-frequency contours. \textbf{b}, the comparisons of Hamiltonian (left) and measured (right) iso-frequency contours. In the analytical iso-frequency contours, TM-like and TE-like bands are again shown in red and blue, respectively, and they use the same color bars as in \textbf{a}. In the measured iso-frequency contours, both TM-like and TE-like resonances are imaged together, with the color bar indicating the total transmitted intensity received by the CCD. The dotted line in the first measurement of the iso-frequency contour indicates the first moiré Briouillin zone. Experimental images are processed using distortion correction and high dynamic range exposure (see Supplementary Material Section 6). 
}
\label{fig3}
\end{figure*}

\begin{figure*}[ht]
\centering
\includegraphics[width=17cm]{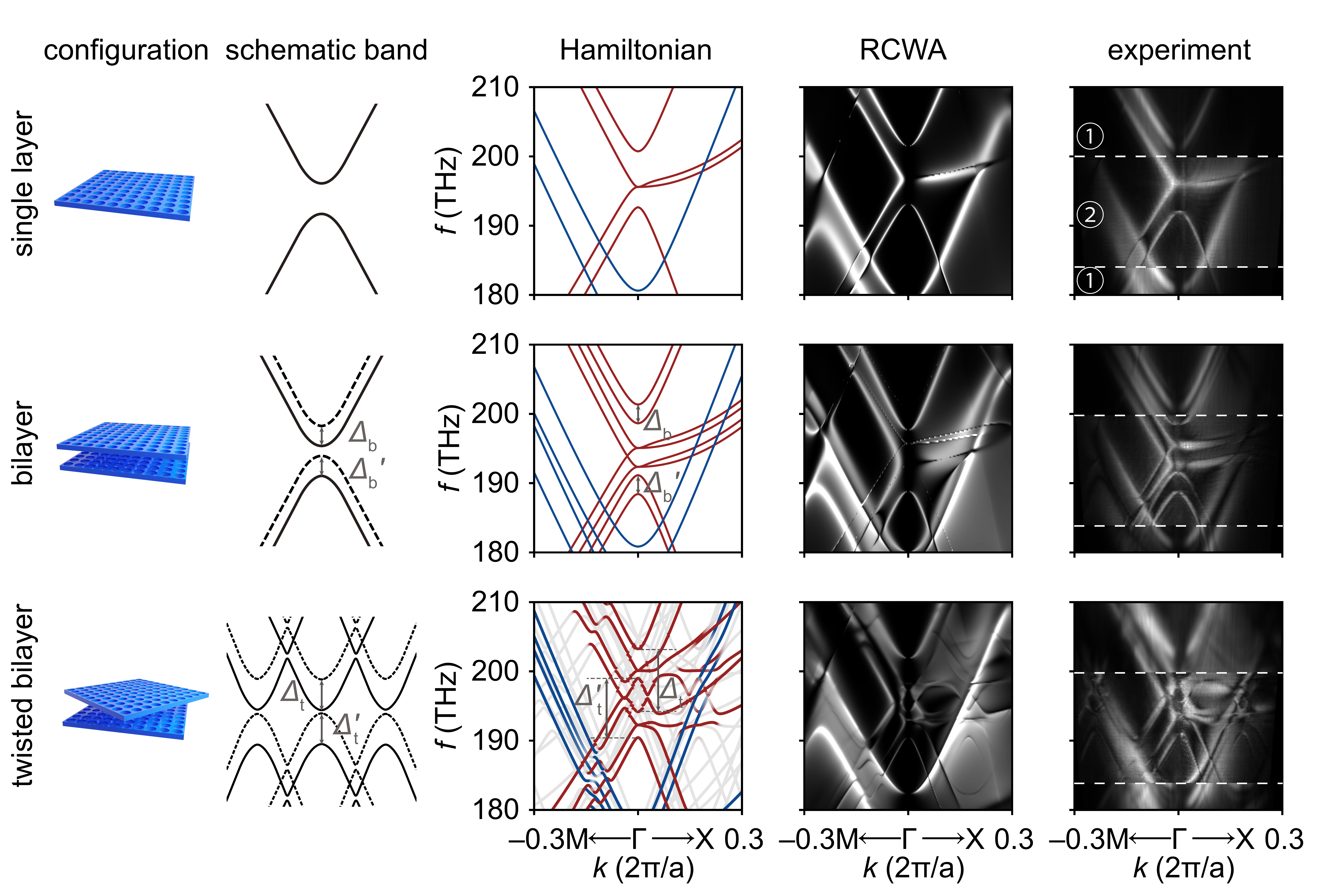}
\caption{\textbf{Schematic, analytical, first-principle, and measured band structures of single-layer, bilayer, and twisted bilayer photonic crystal slabs with a twist angle of  $\theta=10.0\degree$. First column}, the photonic crystal slab configurations. \textbf{Second column}, the schematic band structures. \textbf{Third column}, TM-like (red) and TE-like (blue) Hamiltonian band structures. The gray lines are eigenmodes that are invisible in the measurement. \textbf{Fourth column}, rigorous coupled wave analysis (RCWA) calculation.  \textbf{Fifth  column}, experimental measurements. Band structures in Region 1 are measured by the SuperK laser, and band structures in Region 2 are measured by the Santec TSL tunable laser. 
}
\label{fig4}
\end{figure*}

\begin{figure*}[ht]
\centering
\includegraphics[width=17cm]{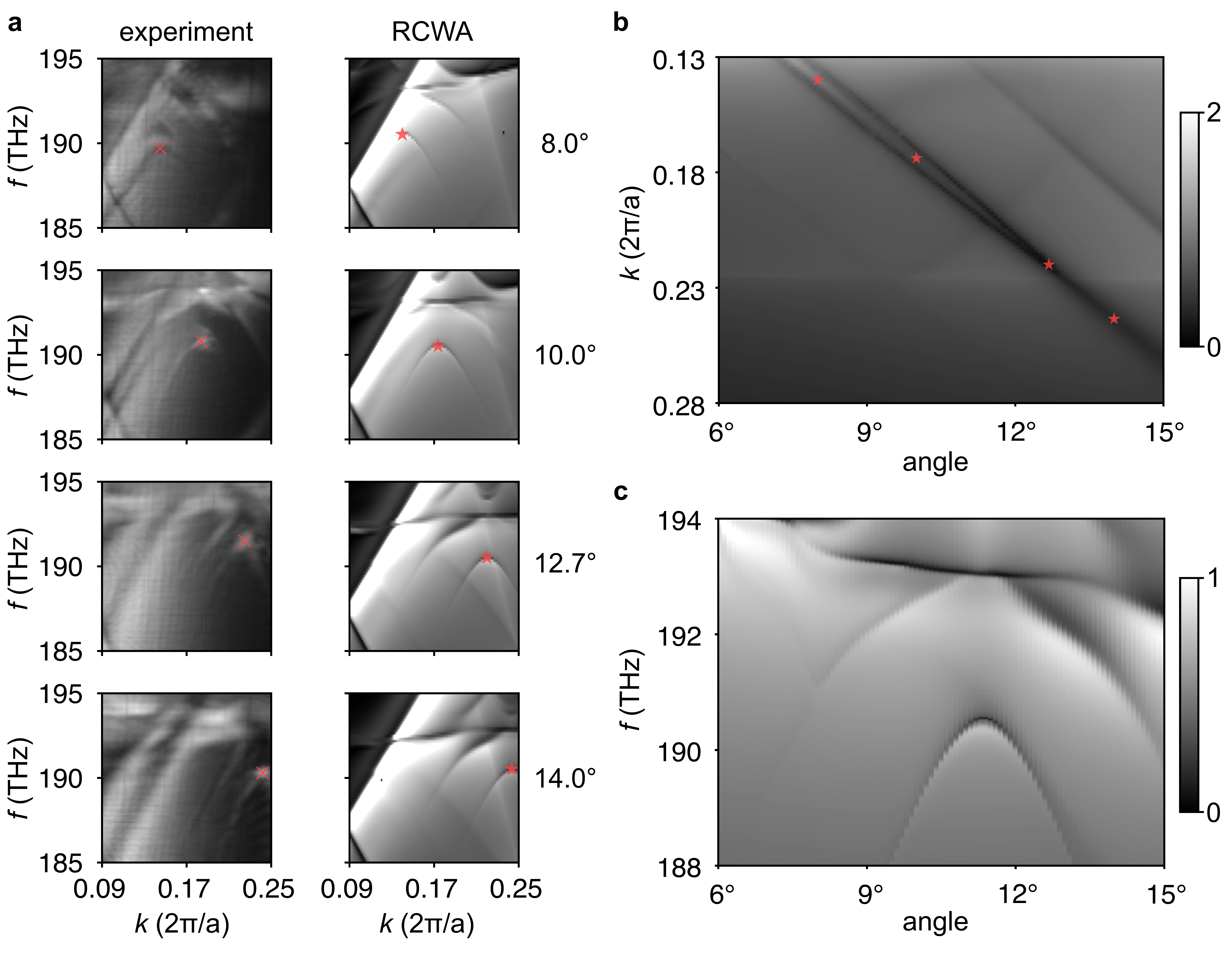}
\caption{\textbf{Angle dependence of optical resonances. a}, The experimental (left column) and RCWA (right column) band structures of twisted bilayer photonic crystal slabs with different twist angles. The band edge is traced by red crosses in the experiment and red stars in the RCWA. \textbf{b}, The wavevectors of the twisted bilayer photonic crystal as a function of twist angle in RCWA, where $f=190.5$ THz \textbf{c}, The resonant frequencies of the twisted bilayer photonic crystal as a function of twist angle in RCWA, where $k=0.189b$. The color bar represents the intensity. 
}
\label{fig5}
\end{figure*}

We start by analyzing the optical scattering properties of a single layer photonic crystal slab and of the twisted bilayer photonic crystal slabs. A photonic crystal slab is a dielectric structure that is finite in the $z$-direction and periodic in $xy$ plane. The periodicity can be represented by a lattice, either in real space or in reciprocal space. For example, a square lattice in the reciprocal space is a set of points $\mathcal{G}_1 \equiv\{( {i} \hat{\mathbf{x}}+ {j} \hat{\mathbf{y}}) 2 \pi / a \mid {i}, {j} \in \mathbb{Z}\}$, where $\hat{\mathbf{x}}$, $\hat{\mathbf{y}}$ are the unit vectors in the $xy$ plane, and $a$ is its period in real space.  When two identical square lattices are twisted against each other (Fig.~\ref{fig1}a), they create moiré lattices in real space (Fig.~\ref{fig1}b). The reciprocal lattice of TBPhC involves a large number of wavevectors due to the scattering by the periodic structure in both layers\cite{Lou2021-dv, Lubin2013-fu} (see Supplementary Material Section 2). The lattice of the twisted layer with a twist angle $\theta$ in the reciprocal space would be $\mathcal{G}_2 \equiv\left\{\left(i \hat{\mathbf{x}}^{\prime}+j \hat{\mathbf{y}}^{\prime}\right) 2 \pi / \theta \mid i, j \in \mathbb{Z}\right\}$, where $\hat{\mathbf{x}}^{\prime}$, $\hat{\mathbf{y}}^{\prime}$ are the unit vectors in the twisted $xy$ plane, namely $\hat{\mathbf{x}}^{\prime}=\cos (\theta) \hat{\mathbf{x}}+\sin (\theta) \hat{\mathbf{y}}$, $\hat{\mathbf{y}}^{\prime}=\cos (\theta) \hat{\mathbf{y}}-\sin (\theta) \hat{\mathbf{x}}$. The reciprocal lattice for the TBPhC is the sum of the reciprocal lattices for each layer, i.e.
\begin{equation}
\begin{aligned}
\mathcal{G}_{\mathrm{TB}} & \equiv\left\{\left(i \hat{\mathbf{x}}+j \hat{\mathbf{y}}+k \hat{\mathbf{x}}^{\prime}+l \hat{\mathbf{y}}^{\prime}\right) 2 \pi / a \mid i, j, k, l \in \mathbb{Z}\right\} \\
& \equiv\left\{\mathbf{G}_{(i, j, k, l)} \mid i, j, k, l \in \mathbb{Z}\right\},
\end{aligned}
\end{equation}
However, only a small subset of wavevectors needs to be considered for light scattering. For TBPhC, the subset of wavevectors of particular interest, which is closely related to the moiré pattern, is
\begin{equation}
\begin{aligned}
\mathcal{G}_{\mathrm{m}} & \equiv\left\{\left({i} \hat{\mathbf{x}}+{j} \hat{\mathbf{y}}-{i} \hat{\mathbf{x}}^{\prime}-{j} \hat{\mathbf{y}}^{\prime}\right) 2 \pi / a \mid {i}, {j} \in \mathbb{Z}\right\}\\
&=\left\{\mathbf{G}_{(i,j,-i,-j)} \mid i, j \in \mathbb{Z}\right\}\\
&\equiv\left\{\mathbf{G}_{\mathrm{m},(i, j)} \mid i, j \in \mathbb{Z}\right\},
\end{aligned}
\end{equation}

where $\mathbf{G}_{\mathrm{m},(i, j)}$ are moiré wavevectors. In particular, the properties of the TBPhC that are strongly dependent on the twist angle arise from a scattering process in which incident light with in-plane wavevector $\mathbf{k}_{\mathrm{inc}}$ is scattered by a moiré wavevector $\mathbf{G}_{ \mathrm{m},(i, j)}$. Such scattering process gives rise to resonances when $\mathbf{k}=\mathbf{k}_{\mathrm{inc}}+\mathbf{G}_{ \mathrm{m},(i, j)}$  matches with the wavevector of the resonant modes of the system (see Supplementary Material Section 1). The first-order moiré wavevectors $\mathbf{G}_{\mathrm{m},(\pm 1,0)}$ or $\mathbf{G}_{\mathrm{m},(0, \pm 1)}$ are determined by the shortest single-layer wavevectors $\mathbf{G}_{\mathbf(\pm 1,0,0,0)}$ and $\mathbf{G}_{(0,0,\pm 1,0)}$ or $\mathbf{G}_{(0, \pm 1,0,0)}$ and $\mathbf{G}_{(0,0,0, \pm 1)}$ (see Fig.~\ref{fig1}c). The next higher-order moiré wavevectors within imaging range $\mathbf{G}_{\mathrm{m},(\pm 1, \pm 1)}$ and $\mathbf{G} _{\mathrm{m},(\pm 1, \mp 1)}$ are determined by $\mathbf{G}_{(\pm 1, \pm 1,0,0)}$ and $\mathbf{G}_{(0,0,\pm 1, \pm 1)}$ (see Supplemental Material Section 2). In our experiment, only the scatterings by a collection of moiré wavevectors $\mathcal{G}_{\mathrm{m}}^\prime=\left\{\mathbf{G}_{\mathrm{m},(\pm 1,0)}, \mathbf{G}_{\mathrm{m},(0, \pm 1)}, \mathbf{G}_{\mathrm{m},(\pm 1, \pm 1)}, \mathbf{G}_{\mathrm{m},(\pm 1, \mp 1)}\right\}$ are observable. Other higher-order moiré wavevector scatterings are negligible when the low-frequency range and small twist angles are considered\cite{Lubin2013-fu, Lou2022-pd}. The first moiré Brillouin zone of the TBPhC slabs is then defined by first-order moiré wavevectors $\mathbf{G}_{\mathrm{m},(\pm 1, 0)}$ and $\mathbf{G}_{\mathrm{m},(0, \pm 1)}$ as shown in Fig.~\ref{fig1}c.

We use two fully suspended nanostructured silicon nitride membranes to fabricate the twisted bilayer photonic crystal slabs (Fig.~\ref{fig2}a). A 150-nm silicon oxide (SiO$_{2}$) layer is first deposited and patterned on the silicon wafer, after which a 439-nm low-stress silicon nitride layer (Si$_{3}$N$_{4}$) is deposited on top of the patterned SiO$_{2}$. A \SI{400}{\micro\metre} $\times$ \SI{400}{\micro\metre} large photonic crystal is patterned into the silicon nitride layer through e-beam lithography: The square lattice photonic crystal has a lattice constant of $a = 1220$ nm and circular air holes with a radius of $r = 502$ nm. The photonic crystal is shallowly etched for an etch depth of $t_{\mathrm{PhC}}=320$ nm. To suspend the photonic crystal slab, the silicon wafer layer underneath is fully etched away from the back side until the etching reaches the SiO$_{2}$ layer. The SiO$_{2}$ layer is then removed through wet etching. A spacing and adhesion layer (SU8 2002) is patterned on top of the Si$_{3}$N$_{4}$ layer in one wafer. Finally, two wafers are bonded together through flip chip bonding techniques that control the relative angle within a precision of $0.1\degree$. There is a $550\pm50$ nm air gap between the two photonic crystal membranes. A moiré pattern appears in the microscope image after the flip chip bonding. (see Supplementary Material Section 3-4). 

We measure the momentum-space-resolved optical response of the sample through a free-space band structure measurement setup\cite{Jin2019-lo, Yin2020-ze}(see Fig.~\ref{fig2}b).  Two lasers are used for the frequency domain and momentum space measurements: the SuperK continuum laser with SuperK SELECT filter ranging from 1100 nm – 1700 nm with a frequency resolution of 6.4 nm – 19.8 nm, and the TSL-550 Santec tunable laser ranging from 1500 nm – 1630 nm with a frequency resolution of 0.003 nm. The laser is sent through the first polarizer (P1) before the light enters the back focal plane of an infinity-corrected objective lens (O1). The incident light focuses on the sample plane and excites the eigenmodes of TBPhC slabs at the same wavelength. These eigenmodes are scattered by the reciprocal lattices, which creates resonances with finite linewidths, as well as by fabrication disorder\cite{Joannopoulos2008-eq}, which further broadens the linewidth of the resonances. A confocal objective lens (O2) collects transmitted light and projects it into momentum space onto its back focal plane. After passing through a 4-$f$ system with second polarizer (P2) that is cross-polarized with P1, the momentum space image is magnified 1.33 times and imaged on a monochromatic CCD camera. P2 is used to block directly transmitted light from P1 and pass TBPhC’s resonances with different polarization states (see Supplementary Material Section 5). This imaging system provides a $k$-space resolution of $1.48\times 10^4$ m$^{-1}$/pixel. The $k$-space imaging range and resolution, in units of the Brillouin zone, depend on the numerical aperture of the objective lens, the incident wavelength, and the lattice constant. For example, when the incident wavelength is 1550 nm and the single-layer PhC lattice constant is $a=1220$ nm, the $k$-space imaging range in length covers 0.53$b$, and the resolution is 0.003$b$/pixel, where $b=2\pi/a$ is the single-layer reciprocal lattice constant. As for the $10.0\degree$ twist bilayer structure, the $k$-space imaging range in length covers 5.6$b_{\mathrm{m}}$, and the resolution is 0.02$b_{\mathrm{m}}$/pixel, where $b_{\mathrm{m}}=2\pi/a_{\mathrm{m}}=4\pi \sin(\theta/2)/a$ is the moiré reciprocal lattice constant (see Supplementary Material Section 6).  

Analytically, the resonant frequencies of the TBPhC can be derived as the eigenfrequencies of a Hamiltonian\cite{Joannopoulos2008-eq, Sakoda2004-xl}  that describes various slab modes and their interactions. We set the unperturbed uniform slab modes (i.e., waveguide modes) as bases and treat the lattice-facilitated scattering as coupling terms in the Hamiltonian matrix. For an incident beam of light with in-plane wavevector $\mathbf{k}_{\text{inc}}$, the relevant slab modes that need to be considered as the bases for the Hamiltonian can be denoted as $\ket{\mathbf{G}_{(i,j,k,l)}}_{{h}}$, or $\ket{{i,j,k,l}}_{{h}}$ for short, with the corresponding field distribution:
\begin{equation}
\mathbf{E}(\mathbf{r}) \equiv \phi_h(z) e^{-i\left(\mathbf{G}_{(i, j, k, l)}+\mathbf{k}_{\text{inc}}\right) \cdot \mathbf{r}_{\|}}
\end{equation}

where $h = 1, 2$ denotes which layer the mode is localized in and $\phi_h(z)$ is the mode profile in the $z$ direction for the frequency range of interest. Many of these bases have a strongly angle-dependent resonant frequency, which follows from the angle dependence of their in-plane wavevectors. As a result, the eigenfrequencies of the Hamiltonian matrix also have a strong twist angle dependence. Examples of the theoretical band structures are shown in Fig.~\ref{fig3}a. Because the first-order moiré scattering is dominant, to construct the Hamiltonian, we choose a set of bases that includes those with wavevectors from individual layer’s reciprocal lattice, such as $\ket{\mathbf{G}_{(\pm 1, 0 , 0, 0)}}_{{h}}$  and $\ket{\mathbf{G}_{(0, \pm 1 , 0, 0)}}_{{h}}$, as well as first-order moiré bases, such as $\ket{\mathbf{G}_{\mathrm{m},(\pm 1,0)}}_{h}$ and $\ket{\mathbf{G}_{\mathrm{m},(0,\pm 1)}}_{h}$, where $h \in\{1,2\}$ labels which layer the mode is localized in (see Supplementary Material Section 7). However, the measurement, which is the frequency domain optical response, does not reflect all the eigenfrequencies in the Hamiltonian calculation. The eigenmodes with more $\ket{\mathbf{G}_{(i,j, 0,0)}}_1$ components in the first layer have stronger coupling strength to the input light. Similarly, eigenmodes with more $\ket{\mathbf{G}_{(0,0, k,l)}}_2$ components in the second layer have stronger coupling strengths to the output light. Eigenmodes with fewer components of $\ket{\mathbf{G}_{(i,j, 0,0)}}_1$ and $\ket{\mathbf{G}_{(0,0, k,l)}}_2$ are expected to manifest less in the transmission spectrum (See Supplementary Material Section 7). In Fig.~\ref{fig3}a, the color bars of the TM-like and TE-like modes represent the coupling strength. Eigenmodes in semi-transparent color are invisible to the measurement. In all following TBPhC band structures, the $x$-axis is scaled by the single-layer photonic crystal reciprocal lattice constant, while the directions of $\Gamma \rightarrow \mathrm{M}$ and $\Gamma \rightarrow \mathrm{X}$ refer to the high symmetry axis of the moiré lattice. The experimental TBPhC iso-frequency contours show good quantitative agreement with the Hamiltonian results (Fig.~\ref{fig3}b). Their frequencies ($f_1=189.7$ THz, $f_2=192.7$ THz, $f_3=194.7$ THz, $f_4 =201.2$ THz) are shown as dashed lines in the analytical band structure in Fig.~\ref{fig3}a. The dotted lines in the first experiment iso-frequency contour in Fig.~\ref{fig3}b indicate moiré lattice. The iso-frequency contour in the first moiré Brillouin zone repeats itself by a translation through the first-order moiré vectors $\mathbf{G}_{\mathrm{m},(\pm 1,0)}$ and $ \mathbf{G}_{\mathrm{m},(0, \pm 1)}$. Similar but much weaker repetitions of iso-frequency contours also appear through the translation of $\mathbf{G}_{\mathrm{m},(\pm 1, \pm 1)}$ and $\mathbf{G} _{\mathrm{m},(\pm 1, \mp 1)}$, indicating a reduced scattering intensity of this order (see Supplementary Material section 2). The repeated iso-frequency contours are not perfectly identical, which coincides with the quasicrystalline feature of the moiré lattice (see Supplementary Material section 1).

The observed band structures reflect band-folding and band-hybridization of single-layer photonic crystal bands. To model the observed band structure of the TBPhC slabs, we use a rigorous coupled-wave analysis (RCWA)\cite{Liu2012-uc}. In contrast to the Hamiltonian approach, RCWA seeks to describe the system from first principles using Maxwell’s equations, and the result of RCWA contains more information, including resonance line width and intensity. We compare Hamiltonian calculations (third column),  RCWA calculations (fourth column), and experimental measurements (last column) of the single-layer photonic crystal slab (first row), bilayer photonic crystal slabs (second row), and TBPhC slabs at a twist angle of 10.0$\degree$ (last row) in Fig.~\ref{fig4}.  In the second column of Fig.~\ref{fig4},  we illustrate the behavior of the TM-like parabolic dispersive bands, which is a prominent feature in the band structure of this system. In the bilayer photonic crystal slabs where the twist angle $\theta = 0 \degree$, the parabolic bands are doubled into two sets vertically (solid$\rightarrow$dashed) as a result of interlayer coupling and band hybridization. In the twisted bilayer photonic crystal slabs, bands are folded back to the first moiré Brillouin zone\cite{Lou2021-dv}, the second moiré Brillouin zone can be visualized in the field-of-view of our setup. The frequency of bands also changes due to the band hybridization\cite{Tang2021-bx, Carr2017-lh} caused by the interlayer coupling. More features can be visualized in the following three columns. In the Hamiltonian, both the TM-like (red) and TE-like (blue) modes are plotted, and the color bars indicating the coupling strength are the same as Fig~\ref{fig3}. In the RCWA calculations, the cross-polarization filter was applied to remove the transmission background. In the measurement, region 1 is measured by the SuperK select laser, and the resolution is lower; region 2 is measured by the Santec TSL tunable laser, and the resolution is higher. The following results were observed from Hamiltonian, RCWA, and measurement results: For the single-layer PhC slab, there are four TM-like bands (red lines). For the bilayer photonic crystal slab, four TM-like bands split into two sets. The gap between each set is $\Delta_{\mathrm{b}}=2.2$ THz for the upper parabolic bands and $\Delta_{\mathrm{b}}^\prime=2.8$ THz for the bottom parabolic bands, which corresponds to the coupling strength in band hybridizations\cite{Tang2021-bx, Carr2017-lh}. Notice that the gap of the perfectly aligned bilayer photonic slab and the misaligned bilayer photonic slab are the same (see Supplementary Material Section 8).  For the TBPhC slab, eight more TM-like bands emerge at $\mathbf{G}_{\mathrm{m}(0, –1)}$, which originated from two sets of TM-like bands in the bilayer PhC slab. The gap is $\Delta_{\mathrm{t}}=7.0$ THz for the upper parabolic bands and $\Delta_{\mathrm{t}}^\prime=7.3$ THz for the bottom parabolic bands. The gap is changed in the TBPhC slab because of the change of the interlayer coupling strength, which is caused by the lattice mismatch. The error between RCWA and measurement result is $\pm 1$ THz at $\Gamma$-point (see Supplementary Material Section 9). Due to the fabrication disorder, the measurement band structures have wider line widths compared to the RCWA. From this comparison between Hamiltonian, RCWA, and experiment, we see a high degree of consistency between the analytical, numerical, and experimental results, which verifies our understanding of the origin of the guided resonances in this system.

One major attraction of TBPhC lies in its twist-angle-tunable guided resonances. Resonances in TBPhC are typically a mixture of angle-independent and angle-dependent resonances. The angle-dependent resonances are strongly associated with the moiré wavevector $\mathbf{G}_{\mathrm{m}}$ and therefore the twist angle\cite{Lou2022-fp, Lou2022-pd}. In Fig.~\ref{fig5}, we compared the measured and calculated band structures of twisted bilayer photonic crystals with different twist angles 8.0$\degree$, 10.0$\degree$, 12.7$\degree$, and 14.0$\degree$(Fig.~\ref{fig5}a). The measured iso-frequency contours for these four angles are in Supplementary Material Section 10. As illustrated in Fig.~\ref{fig5}, the twisted bilayer structure has a set of parabolic bands centering the moire wavevector $\mathbf{G}_{\mathrm{m,(1,0)}}$. As the angle gets larger, the magnitude of the moiré wavevector $\mathbf{G}_{\mathrm{m}}$ becomes larger, and a shift of the whole band towards larger $k$ is observed. The band edge is marked by red crosses and stars in Fig.~\ref{fig5}a for clarity. To better see how the band edge moves as the twist angle is varied, we plot the transmission as a function of wavevector and twist angle for a fixed frequency of $190.5$THz in Fig.~\ref{fig5}b, where the points marked by red stars correspond to the same parameters as those marked in Fig.~\ref{fig5}a. In this parameter range, the tunable resonance can be well explained by the band folding picture, where the parabolic bands are moved according to the corresponding moire wavevector.  We also provide the angle dependence of the resonance frequency at $k=0.189b$ (Fig.~\ref{fig5}c), where the transmission dips around frequency of $190.3$THz also follow a parabolic shape. For higher frequencies, more complicated twist angle dependence can be observed. The angle dependency of guided resonances helps identify the tunability of scattering properties in the twisted bilayer photonic crystal slabs and foresee the potential of next-generation bilayer reconfigurable devices such as tunable filters and beam steering.

In conclusion, we show in this paper, how the twist angle between two photonic crystal slabs can be used to tune the optical band structures of the assembly. We built twisted bilayer photonic crystal structures through nanofabrication and developed a fundamental understanding of their complex, unconventional optical properties. Specifically, we theoretically demonstrated and experimentally measured the band structure of twisted bilayer photonic crystal slabs in the optical frequency range. In particular, we especially analyzed the first-order moiré scattering behavior that is observable from iso-frequency contours along with the band hybridization and band-folding behavior from the band structure. Our work establishes the basis for engineering electromagnetic wave propagation in twisted bilayer photonic structures, generating a new suite of optical properties through the creation of synthetic moiré systems controllable by twist angle and interlayer coupling. The band structure tunability provides a starting point for the understanding of other optical properties in the dielectric twisted bilayer systems, including bound-state-in-continuum, quasicrystalline optics, chirality, polarimetry, nontrivial-topological modes, superscattering, etc. The fabrication, measurement, and analysis approaches presented here will also be a platform for building complex bilayer nanomaterials and controlling electromagnetic waves with mechanical reconfigurability, such as interlayer gap, twist angle, and sliding distance. Our experiment also serves as a foundation for developing bilayer flat-optical devices such as tunable filters, tunable lasers, adaptive sensors, and LiDAR.

\bibliography{ref.bib}

\begin{thebibliography}{10}
\expandafter\ifx\csname url\endcsname\relax
  \def\url#1{\texttt{#1}}\fi
\expandafter\ifx\csname urlprefix\endcsname\relax\def\urlprefix{URL }\fi
\providecommand{\bibinfo}[2]{#2}
\providecommand{\eprint}[2][]{\url{#2}}

\bibitem{Wang2020-ce}
\bibinfo{author}{Wang, P.}, \bibinfo{author}{Zheng, Y.}, \bibinfo{author}{Chen,
  X.}, \bibinfo{author}{Huang, C.}, \bibinfo{author}{Kartashov, Y.~V.},
  \bibinfo{author}{Torner, L.}, \bibinfo{author}{Konotop, V.~V.} \&
  \bibinfo{author}{Ye, F.}
\newblock \bibinfo{title}{Localization and delocalization of light in photonic
  moir{\'e} lattices}.
\newblock \emph{\bibinfo{journal}{Nature}} \textbf{\bibinfo{volume}{577}},
  \bibinfo{pages}{42--46} (\bibinfo{year}{2020}).

\bibitem{Fu2020-zh}
\bibinfo{author}{Fu, Q.}, \bibinfo{author}{Wang, P.}, \bibinfo{author}{Huang,
  C.}, \bibinfo{author}{Kartashov, Y.~V.}, \bibinfo{author}{Torner, L.},
  \bibinfo{author}{Konotop, V.~V.} \& \bibinfo{author}{Ye, F.}
\newblock \bibinfo{title}{Optical soliton formation controlled by angle
  twisting in photonic moir{\'e} lattices}.
\newblock \emph{\bibinfo{journal}{Nat. Photonics}}
  \textbf{\bibinfo{volume}{14}}, \bibinfo{pages}{663--668}
  (\bibinfo{year}{2020}).

\bibitem{Talukdar2022-jf}
\bibinfo{author}{Talukdar, T.~H.}, \bibinfo{author}{Hardison, A.~L.} \&
  \bibinfo{author}{Ryckman, J.~D.}
\newblock \bibinfo{title}{Moir{\'e} effects in silicon photonic nanowires}.
\newblock \emph{\bibinfo{journal}{ACS Photonics}}  (\bibinfo{year}{2022}).

\bibitem{Mao2021-ye}
\bibinfo{author}{Mao, X.-R.}, \bibinfo{author}{Shao, Z.-K.},
  \bibinfo{author}{Luan, H.-Y.}, \bibinfo{author}{Wang, S.-L.} \&
  \bibinfo{author}{Ma, R.-M.}
\newblock \bibinfo{title}{Magic-angle lasers in nanostructured moir{\'e}
  superlattice}.
\newblock \emph{\bibinfo{journal}{Nat. Nanotechnol.}}
  \textbf{\bibinfo{volume}{16}}, \bibinfo{pages}{1099--1105}
  (\bibinfo{year}{2021}).

\bibitem{Zhou2020-ox}
\bibinfo{author}{Zhou, X.}, \bibinfo{author}{Lin, Z.}, \bibinfo{author}{Lu,
  W.}, \bibinfo{author}{Lai, Y.}, \bibinfo{author}{Hou, B.} \&
  \bibinfo{author}{Jiang, J.}
\newblock \bibinfo{title}{Photonic crystals: Twisted quadrupole topological
  photonic crystals (laser photonics rev. 14(8)/2020)} (\bibinfo{year}{2020}).

\bibitem{Shang2021-kh}
\bibinfo{author}{Shang, C.}, \bibinfo{author}{Lu, C.}, \bibinfo{author}{Tang,
  S.}, \bibinfo{author}{Gao, Y.} \& \bibinfo{author}{Wen, Z.}
\newblock \bibinfo{title}{Generation of gradient photonic moir{\'e} lattice
  fields}.
\newblock \emph{\bibinfo{journal}{Opt. Express}} \textbf{\bibinfo{volume}{29}},
  \bibinfo{pages}{29116--29127} (\bibinfo{year}{2021}).

\bibitem{Lin2022-cz}
\bibinfo{author}{Lin, H.-M.}, \bibinfo{author}{Lu, Y.-H.},
  \bibinfo{author}{Chang, Y.-J.}, \bibinfo{author}{Yang, Y.-Y.} \&
  \bibinfo{author}{Jin, X.-M.}
\newblock \bibinfo{title}{Direct observation of a localized flat-band state in
  a mapped moir{\'e} hubbard photonic lattice}.
\newblock \emph{\bibinfo{journal}{Phys. Rev. Appl.}}
  \textbf{\bibinfo{volume}{18}} (\bibinfo{year}{2022}).

\bibitem{Han2015-ak}
\bibinfo{author}{Han, J.-H.}, \bibinfo{author}{Kim, I.}, \bibinfo{author}{Ryu,
  J.-W.}, \bibinfo{author}{Kim, J.}, \bibinfo{author}{Cho, J.-H.},
  \bibinfo{author}{Yim, G.-S.}, \bibinfo{author}{Park, H.-S.},
  \bibinfo{author}{Min, B.} \& \bibinfo{author}{Choi, M.}
\newblock \bibinfo{title}{Rotationally reconfigurable metamaterials based on
  moir{\'e} phenomenon}.
\newblock \emph{\bibinfo{journal}{Opt. Express}} \textbf{\bibinfo{volume}{23}},
  \bibinfo{pages}{17443--17449} (\bibinfo{year}{2015}).

\bibitem{Asboth2016-cd}
\bibinfo{author}{Asb{\'o}th, J.~K.}, \bibinfo{author}{Oroszl{\'a}ny, L.} \&
  \bibinfo{author}{P{\'a}lyi, A.~P.}
\newblock \emph{\bibinfo{title}{A Short Course on Topological Insulators: Band
  Structure and Edge States in One and Two Dimensions}}
  (\bibinfo{publisher}{Springer}, \bibinfo{year}{2016}).

\bibitem{Zeng2021-po}
\bibinfo{author}{Zeng, J.}, \bibinfo{author}{Hu, Y.}, \bibinfo{author}{Zhang,
  X.}, \bibinfo{author}{Fu, S.}, \bibinfo{author}{Yin, H.},
  \bibinfo{author}{Li, Z.} \& \bibinfo{author}{Chen, Z.}
\newblock \bibinfo{title}{Localization-to-delocalization transition of light in
  frequency-tuned photonic moir{\'e} lattices}.
\newblock \emph{\bibinfo{journal}{Opt. Express}} \textbf{\bibinfo{volume}{29}},
  \bibinfo{pages}{25388--25398} (\bibinfo{year}{2021}).

\bibitem{Alnasser2021-nq}
\bibinfo{author}{Alnasser, K.}, \bibinfo{author}{Kamau, S.},
  \bibinfo{author}{Hurley, N.}, \bibinfo{author}{Cui, J.} \&
  \bibinfo{author}{Lin, Y.}
\newblock \bibinfo{title}{Resonance modes in moir{\'e} photonic patterns for
  twistoptics}.
\newblock \emph{\bibinfo{journal}{OSA Continuum}} \textbf{\bibinfo{volume}{4}},
  \bibinfo{pages}{1339} (\bibinfo{year}{2021}).

\bibitem{Kartashov2021-kl}
\bibinfo{author}{Kartashov, Y.~V.}, \bibinfo{author}{Ye, F.},
  \bibinfo{author}{Konotop, V.~V.} \& \bibinfo{author}{Torner, L.}
\newblock \bibinfo{title}{Multifrequency solitons in
  {Commensurate-Incommensurate} photonic moir{\'e} lattices}.
\newblock \emph{\bibinfo{journal}{Phys. Rev. Lett.}}
  \textbf{\bibinfo{volume}{127}}, \bibinfo{pages}{163902}
  (\bibinfo{year}{2021}).

\bibitem{Khurgin2000-au}
\bibinfo{author}{Khurgin, J.~B.}
\newblock \bibinfo{title}{Light slowing down in moir{\'e} fiber gratings and
  its implications for nonlinear optics} (\bibinfo{year}{2000}).

\bibitem{Beravat2016-vd}
\bibinfo{author}{Beravat, R.}, \bibinfo{author}{Wong, G. K.~L.},
  \bibinfo{author}{Frosz, M.~H.}, \bibinfo{author}{Xi, X.~M.} \&
  \bibinfo{author}{Russell, P. S.~J.}
\newblock \bibinfo{title}{Twist-induced guidance in coreless photonic crystal
  fiber: A helical channel for light}.
\newblock \emph{\bibinfo{journal}{Sci Adv}} \textbf{\bibinfo{volume}{2}},
  \bibinfo{pages}{e1601421} (\bibinfo{year}{2016}).

\bibitem{Wong2012-kx}
\bibinfo{author}{Wong, G. K.~L.}, \bibinfo{author}{Kang, M.~S.},
  \bibinfo{author}{Lee, H.~W.}, \bibinfo{author}{Biancalana, F.},
  \bibinfo{author}{Conti, C.}, \bibinfo{author}{Weiss, T.} \&
  \bibinfo{author}{Russell, P. S.~J.}
\newblock \bibinfo{title}{Excitation of orbital angular momentum resonances in
  helically twisted photonic crystal fiber}.
\newblock \emph{\bibinfo{journal}{Science}} \textbf{\bibinfo{volume}{337}},
  \bibinfo{pages}{446--449} (\bibinfo{year}{2012}).

\bibitem{Lou2022-fp}
\bibinfo{author}{Lou, B.}, \bibinfo{author}{Wang, B.},
  \bibinfo{author}{Rodr{\'\i}guez, J.~A.}, \bibinfo{author}{Cappelli, M.} \&
  \bibinfo{author}{Fan, S.}
\newblock \bibinfo{title}{Tunable guided resonance in twisted bilayer photonic
  crystal}.
\newblock \emph{\bibinfo{journal}{Sci Adv}} \textbf{\bibinfo{volume}{8}},
  \bibinfo{pages}{eadd4339} (\bibinfo{year}{2022}).

\bibitem{Lou2021-dv}
\bibinfo{author}{Lou, B.}, \bibinfo{author}{Zhao, N.}, \bibinfo{author}{Minkov,
  M.}, \bibinfo{author}{Guo, C.}, \bibinfo{author}{Orenstein, M.} \&
  \bibinfo{author}{Fan, S.}
\newblock \bibinfo{title}{Theory for twisted bilayer photonic crystal slabs}.
\newblock \emph{\bibinfo{journal}{Phys. Rev. Lett.}}
  \textbf{\bibinfo{volume}{126}}, \bibinfo{pages}{136101}
  (\bibinfo{year}{2021}).

\bibitem{https://doi.org/10.48550/arxiv.2211.07230}
\bibinfo{author}{Nguyen, D. H.~M.}, \bibinfo{author}{Devescovi, C.},
  \bibinfo{author}{Nguyen, D.~X.}, \bibinfo{author}{Nguyen, H.~S.} \&
  \bibinfo{author}{Bercioux, D.}
\newblock \bibinfo{title}{Fermi arc reconstruction in synthetic photonic
  lattice} (\bibinfo{year}{2022}).
\newblock \urlprefix\url{https://arxiv.org/abs/2211.07230}.

\bibitem{Wang2022-nf}
\bibinfo{author}{Wang, H.}, \bibinfo{author}{Ma, S.}, \bibinfo{author}{Zhang,
  S.} \& \bibinfo{author}{Lei, D.}
\newblock \bibinfo{title}{Intrinsic superflat bands in general twisted bilayer
  systems}.
\newblock \emph{\bibinfo{journal}{Light Sci Appl}}
  \textbf{\bibinfo{volume}{11}}, \bibinfo{pages}{159} (\bibinfo{year}{2022}).

\bibitem{Dong2021-mh}
\bibinfo{author}{Dong, K.}, \bibinfo{author}{Zhang, T.}, \bibinfo{author}{Li,
  J.}, \bibinfo{author}{Wang, Q.}, \bibinfo{author}{Yang, F.},
  \bibinfo{author}{Rho, Y.}, \bibinfo{author}{Wang, D.},
  \bibinfo{author}{Grigoropoulos, C.~P.}, \bibinfo{author}{Wu, J.} \&
  \bibinfo{author}{Yao, J.}
\newblock \bibinfo{title}{Flat bands in magic-angle bilayer photonic crystals
  at small twists}.
\newblock \emph{\bibinfo{journal}{Phys. Rev. Lett.}}
  \textbf{\bibinfo{volume}{126}}, \bibinfo{pages}{223601}
  (\bibinfo{year}{2021}).

\bibitem{Lubin2013-fu}
\bibinfo{author}{Lubin, S.~M.}, \bibinfo{author}{Hryn, A.~J.},
  \bibinfo{author}{Huntington, M.~D.}, \bibinfo{author}{Engel, C.~J.} \&
  \bibinfo{author}{Odom, T.~W.}
\newblock \bibinfo{title}{Quasiperiodic moir{\'e} plasmonic crystals}.
\newblock \emph{\bibinfo{journal}{ACS Nano}} \textbf{\bibinfo{volume}{7}},
  \bibinfo{pages}{11035--11042} (\bibinfo{year}{2013}).

\bibitem{Wu2018-kb}
\bibinfo{author}{Wu, Z.}, \bibinfo{author}{Chen, X.}, \bibinfo{author}{Wang,
  M.}, \bibinfo{author}{Dong, J.} \& \bibinfo{author}{Zheng, Y.}
\newblock \bibinfo{title}{{High-Performance} ultrathin active chiral
  metamaterials}.
\newblock \emph{\bibinfo{journal}{ACS Nano}} \textbf{\bibinfo{volume}{12}},
  \bibinfo{pages}{5030--5041} (\bibinfo{year}{2018}).

\bibitem{Yi2022-nh}
\bibinfo{author}{Yi, C.-H.}, \bibinfo{author}{Park, H.~C.} \&
  \bibinfo{author}{Park, M.~J.}
\newblock \bibinfo{title}{Strong interlayer coupling and stable topological
  flat bands in twisted bilayer photonic moir{\'e} superlattices}.
\newblock \emph{\bibinfo{journal}{Light Sci Appl}}
  \textbf{\bibinfo{volume}{11}}, \bibinfo{pages}{289} (\bibinfo{year}{2022}).

\bibitem{Nguyen2022-bt}
\bibinfo{author}{Nguyen, D.~X.}, \bibinfo{author}{Letartre, X.},
  \bibinfo{author}{Drouard, E.}, \bibinfo{author}{Viktorovitch, P.},
  \bibinfo{author}{Nguyen, H.~C.} \& \bibinfo{author}{Nguyen, H.~S.}
\newblock \bibinfo{title}{Magic configurations in moir{\'e} superlattice of
  bilayer photonic crystals: Almost-perfect flatbands and unconventional
  localization}.
\newblock \emph{\bibinfo{journal}{Phys. Rev. Research}}
  \textbf{\bibinfo{volume}{4}} (\bibinfo{year}{2022}).

\bibitem{Huang2022-yp}
\bibinfo{author}{Huang, L.}, \bibinfo{author}{Zhang, W.} \&
  \bibinfo{author}{Zhang, X.}
\newblock \bibinfo{title}{Moir{\'e} quasibound states in the continuum}.
\newblock \emph{\bibinfo{journal}{Phys. Rev. Lett.}}
  \textbf{\bibinfo{volume}{128}}, \bibinfo{pages}{253901}
  (\bibinfo{year}{2022}).

\bibitem{Hu2021-ty}
\bibinfo{author}{Hu, G.}, \bibinfo{author}{Zheng, C.}, \bibinfo{author}{Ni,
  J.}, \bibinfo{author}{Qiu, C.-W.} \& \bibinfo{author}{Al{\`u}, A.}
\newblock \bibinfo{title}{Enhanced light-matter interactions at photonic
  magic-angle topological transitions}.
\newblock \emph{\bibinfo{journal}{Appl. Phys. Lett.}}
  \textbf{\bibinfo{volume}{118}}, \bibinfo{pages}{211101}
  (\bibinfo{year}{2021}).

\bibitem{Sunku2018-uf}
\bibinfo{author}{Sunku, S.~S.}, \bibinfo{author}{Ni, G.~X.},
  \bibinfo{author}{Jiang, B.~Y.}, \bibinfo{author}{Yoo, H.},
  \bibinfo{author}{Sternbach, A.}, \bibinfo{author}{McLeod, A.~S.},
  \bibinfo{author}{Stauber, T.}, \bibinfo{author}{Xiong, L.},
  \bibinfo{author}{Taniguchi, T.}, \bibinfo{author}{Watanabe, K.},
  \bibinfo{author}{Kim, P.}, \bibinfo{author}{Fogler, M.~M.} \&
  \bibinfo{author}{Basov, D.~N.}
\newblock \bibinfo{title}{Photonic crystals for nano-light in moir{\'e}
  graphene superlattices}.
\newblock \emph{\bibinfo{journal}{Science}} \textbf{\bibinfo{volume}{362}},
  \bibinfo{pages}{1153--1156} (\bibinfo{year}{2018}).

\bibitem{Duan2020-yw}
\bibinfo{author}{Duan, J.}, \bibinfo{author}{Capote-Robayna, N.},
  \bibinfo{author}{Taboada-Guti{\'e}rrez, J.},
  \bibinfo{author}{{\'A}lvarez-P{\'e}rez, G.}, \bibinfo{author}{Prieto, I.},
  \bibinfo{author}{Mart{\'\i}n-S{\'a}nchez, J.}, \bibinfo{author}{Nikitin,
  A.~Y.} \& \bibinfo{author}{Alonso-Gonz{\'a}lez, P.}
\newblock \bibinfo{title}{Twisted {Nano-Optics}: Manipulating light at the
  nanoscale with twisted phonon polaritonic slabs}.
\newblock \emph{\bibinfo{journal}{Nano Lett.}} \textbf{\bibinfo{volume}{20}},
  \bibinfo{pages}{5323--5329} (\bibinfo{year}{2020}).

\bibitem{Chen2020-lf}
\bibinfo{author}{Chen, M.}, \bibinfo{author}{Lin, X.}, \bibinfo{author}{Dinh,
  T.~H.}, \bibinfo{author}{Zheng, Z.}, \bibinfo{author}{Shen, J.},
  \bibinfo{author}{Ma, Q.}, \bibinfo{author}{Chen, H.},
  \bibinfo{author}{Jarillo-Herrero, P.} \& \bibinfo{author}{Dai, S.}
\newblock \bibinfo{title}{Configurable phonon polaritons in twisted
  {$\alpha$-MoO3}}.
\newblock \emph{\bibinfo{journal}{Nat. Mater.}} \textbf{\bibinfo{volume}{19}},
  \bibinfo{pages}{1307--1311} (\bibinfo{year}{2020}).

\bibitem{Krasnok2022-bf}
\bibinfo{author}{Krasnok, A.} \& \bibinfo{author}{Al{\`u}, A.}
\newblock \bibinfo{title}{{Low-Symmetry} nanophotonics}.
\newblock \emph{\bibinfo{journal}{ACS Photonics}} \textbf{\bibinfo{volume}{9}},
  \bibinfo{pages}{2--24} (\bibinfo{year}{2022}).

\bibitem{Zhang2021-qg}
\bibinfo{author}{Zhang, Q.}, \bibinfo{author}{Hu, G.}, \bibinfo{author}{Ma,
  W.}, \bibinfo{author}{Li, P.}, \bibinfo{author}{Krasnok, A.},
  \bibinfo{author}{Hillenbrand, R.}, \bibinfo{author}{Al{\`u}, A.} \&
  \bibinfo{author}{Qiu, C.-W.}
\newblock \bibinfo{title}{Interface nano-optics with van der waals polaritons}.
\newblock \emph{\bibinfo{journal}{Nature}} \textbf{\bibinfo{volume}{597}},
  \bibinfo{pages}{187--195} (\bibinfo{year}{2021}).

\bibitem{Hu2021-mk}
\bibinfo{author}{Hu, G.}, \bibinfo{author}{Wang, M.}, \bibinfo{author}{Mazor,
  Y.}, \bibinfo{author}{Qiu, C.-W.} \& \bibinfo{author}{Al{\`u}, A.}
\newblock \bibinfo{title}{Tailoring light with layered and moir{\'e}
  metasurfaces}.
\newblock \emph{\bibinfo{journal}{Trends in Chemistry}}
  \textbf{\bibinfo{volume}{3}}, \bibinfo{pages}{342--358}
  (\bibinfo{year}{2021}).

\bibitem{Chen2021-ef}
\bibinfo{author}{Chen, J.}, \bibinfo{author}{Lin, X.}, \bibinfo{author}{Chen,
  M.}, \bibinfo{author}{Low, T.}, \bibinfo{author}{Chen, H.} \&
  \bibinfo{author}{Dai, S.}
\newblock \bibinfo{title}{A perspective of twisted photonic structures}.
\newblock \emph{\bibinfo{journal}{Appl. Phys. Lett.}}
  \textbf{\bibinfo{volume}{119}}, \bibinfo{pages}{240501}
  (\bibinfo{year}{2021}).

\bibitem{Wu2018-on}
\bibinfo{author}{Wu, Z.}, \bibinfo{author}{Liu, Y.}, \bibinfo{author}{Hill,
  E.~H.} \& \bibinfo{author}{Zheng, Y.}
\newblock \bibinfo{title}{Chiral metamaterials via moir{\'e} stacking}.
\newblock \emph{\bibinfo{journal}{Nanoscale}}  (\bibinfo{year}{2018}).

\bibitem{Liu2022-ru}
\bibinfo{author}{Liu, S.}, \bibinfo{author}{Ma, S.}, \bibinfo{author}{Shao,
  R.}, \bibinfo{author}{Zhang, L.}, \bibinfo{author}{Yan, T.},
  \bibinfo{author}{Ma, Q.}, \bibinfo{author}{Zhang, S.} \&
  \bibinfo{author}{Cui, T.~J.}
\newblock \bibinfo{title}{Moir{\'e} metasurfaces for dynamic beamforming}.
\newblock \emph{\bibinfo{journal}{Sci Adv}} \textbf{\bibinfo{volume}{8}},
  \bibinfo{pages}{eabo1511} (\bibinfo{year}{2022}).

\bibitem{Yao2021-ma}
\bibinfo{author}{Yao, K.}, \bibinfo{author}{Finney, N.~R.},
  \bibinfo{author}{Zhang, J.}, \bibinfo{author}{Moore, S.~L.},
  \bibinfo{author}{Xian, L.}, \bibinfo{author}{Tancogne-Dejean, N.},
  \bibinfo{author}{Liu, F.}, \bibinfo{author}{Ardelean, J.},
  \bibinfo{author}{Xu, X.}, \bibinfo{author}{Halbertal, D.},
  \bibinfo{author}{Watanabe, K.}, \bibinfo{author}{Taniguchi, T.},
  \bibinfo{author}{Ochoa, H.}, \bibinfo{author}{Asenjo-Garcia, A.},
  \bibinfo{author}{Zhu, X.}, \bibinfo{author}{Basov, D.~N.},
  \bibinfo{author}{Rubio, A.}, \bibinfo{author}{Dean, C.~R.},
  \bibinfo{author}{Hone, J.} \& \bibinfo{author}{Schuck, P.~J.}
\newblock \bibinfo{title}{Enhanced tunable second harmonic generation from
  twistable interfaces and vertical superlattices in boron nitride
  homostructures}.
\newblock \emph{\bibinfo{journal}{Sci Adv}} \textbf{\bibinfo{volume}{7}}
  (\bibinfo{year}{2021}).

\bibitem{Hu2020-lh}
\bibinfo{author}{Hu, G.}, \bibinfo{author}{Krasnok, A.},
  \bibinfo{author}{Mazor, Y.}, \bibinfo{author}{Qiu, C.-W.} \&
  \bibinfo{author}{Al{\`u}, A.}
\newblock \bibinfo{title}{Moir{\'e} hyperbolic metasurfaces}.
\newblock \emph{\bibinfo{journal}{Nano Lett.}} \textbf{\bibinfo{volume}{20}},
  \bibinfo{pages}{3217--3224} (\bibinfo{year}{2020}).

\bibitem{Joannopoulos2008-eq}
\bibinfo{author}{Joannopoulos, J.~D.}, \bibinfo{author}{Johnson, S.~G.},
  \bibinfo{author}{Winn, J.~N.} \& \bibinfo{author}{Meade, R.~D.}
\newblock \emph{\bibinfo{title}{Photonic Crystals: Molding the Flow of Light
  (Second Edition)}} (\bibinfo{publisher}{Princeton University Press},
  \bibinfo{year}{2008}).

\bibitem{Tang2021-bx}
\bibinfo{author}{Tang, H.}, \bibinfo{author}{Du, F.}, \bibinfo{author}{Carr,
  S.}, \bibinfo{author}{DeVault, C.}, \bibinfo{author}{Mello, O.} \&
  \bibinfo{author}{Mazur, E.}
\newblock \bibinfo{title}{Modeling the optical properties of twisted bilayer
  photonic crystals}.
\newblock \emph{\bibinfo{journal}{Light Sci Appl}}
  \textbf{\bibinfo{volume}{10}}, \bibinfo{pages}{157} (\bibinfo{year}{2021}).

\bibitem{Tang2022-yh}
\bibinfo{author}{Tang, H.}, \bibinfo{author}{Ni, X.}, \bibinfo{author}{Du, F.},
  \bibinfo{author}{Srikrishna, V.} \& \bibinfo{author}{Mazur, E.}
\newblock \bibinfo{title}{On-chip light trapping in bilayer moir{\'e} photonic
  crystal slabs}.
\newblock \emph{\bibinfo{journal}{Appl. Phys. Lett.}}
  \textbf{\bibinfo{volume}{121}}, \bibinfo{pages}{231702}
  (\bibinfo{year}{2022}).

\bibitem{Lou2022-pd}
\bibinfo{author}{Lou, B.} \& \bibinfo{author}{Fan, S.}
\newblock \bibinfo{title}{Tunable frequency filter based on twisted bilayer
  photonic crystal slabs}.
\newblock \emph{\bibinfo{journal}{ACS Photonics}} \textbf{\bibinfo{volume}{9}},
  \bibinfo{pages}{800--805} (\bibinfo{year}{2022}).

\bibitem{PhysRevLett.128.253901}
\bibinfo{author}{Huang, L.}, \bibinfo{author}{Zhang, W.} \&
  \bibinfo{author}{Zhang, X.}
\newblock \bibinfo{title}{Moir\'e quasibound states in the continuum}.
\newblock \emph{\bibinfo{journal}{Phys. Rev. Lett.}}
  \textbf{\bibinfo{volume}{128}}, \bibinfo{pages}{253901}
  (\bibinfo{year}{2022}).
\newblock
  \urlprefix\url{https://link.aps.org/doi/10.1103/PhysRevLett.128.253901}.

\bibitem{Gan2012-ne}
\bibinfo{author}{Gan, X.}, \bibinfo{author}{Pervez, N.},
  \bibinfo{author}{Kymissis, I.}, \bibinfo{author}{Hatami, F.} \&
  \bibinfo{author}{Englund, D.}
\newblock \bibinfo{title}{A high-resolution spectrometer based on a compact
  planar two dimensional photonic crystal cavity array}.
\newblock \emph{\bibinfo{journal}{Appl. Phys. Lett.}}
  \textbf{\bibinfo{volume}{100}}, \bibinfo{pages}{231104}
  (\bibinfo{year}{2012}).

\bibitem{Wang2014-af}
\bibinfo{author}{Wang, Z.} \& \bibinfo{author}{Yu, Z.}
\newblock \bibinfo{title}{Spectral analysis based on compressive sensing in
  nanophotonic structures}.
\newblock \emph{\bibinfo{journal}{Opt. Express}} \textbf{\bibinfo{volume}{22}},
  \bibinfo{pages}{25608--25614} (\bibinfo{year}{2014}).

\bibitem{Jin2019-lo}
\bibinfo{author}{Jin, J.}, \bibinfo{author}{Yin, X.}, \bibinfo{author}{Ni, L.},
  \bibinfo{author}{Solja{\v c}i{\'c}, M.}, \bibinfo{author}{Zhen, B.} \&
  \bibinfo{author}{Peng, C.}
\newblock \bibinfo{title}{Topologically enabled ultrahigh-q guided resonances
  robust to out-of-plane scattering}.
\newblock \emph{\bibinfo{journal}{Nature}} \textbf{\bibinfo{volume}{574}},
  \bibinfo{pages}{501--504} (\bibinfo{year}{2019}).

\bibitem{Yin2020-ze}
\bibinfo{author}{Yin, X.}, \bibinfo{author}{Jin, J.}, \bibinfo{author}{Solja{\v
  c}i{\'c}, M.}, \bibinfo{author}{Peng, C.} \& \bibinfo{author}{Zhen, B.}
\newblock \bibinfo{title}{Observation of topologically enabled unidirectional
  guided resonances}.
\newblock \emph{\bibinfo{journal}{Nature}} \textbf{\bibinfo{volume}{580}},
  \bibinfo{pages}{467--471} (\bibinfo{year}{2020}).

\bibitem{Sakoda2004-xl}
\bibinfo{author}{Sakoda, K.}
\newblock \emph{\bibinfo{title}{Optical Properties of Photonic Crystals}}
  (\bibinfo{publisher}{Springer Science \& Business Media},
  \bibinfo{year}{2004}).

\bibitem{Liu2012-uc}
\bibinfo{author}{Liu, V.} \& \bibinfo{author}{Fan, S.}
\newblock \bibinfo{title}{{S4} : A free electromagnetic solver for layered
  periodic structures}.
\newblock \emph{\bibinfo{journal}{Comput. Phys. Commun.}}
  \textbf{\bibinfo{volume}{183}}, \bibinfo{pages}{2233--2244}
  (\bibinfo{year}{2012}).

\bibitem{Carr2017-lh}
\bibinfo{author}{Carr, S.}, \bibinfo{author}{Massatt, D.},
  \bibinfo{author}{Fang, S.}, \bibinfo{author}{Cazeaux, P.},
  \bibinfo{author}{Luskin, M.} \& \bibinfo{author}{Kaxiras, E.}
\newblock \bibinfo{title}{Twistronics: Manipulating the electronic properties
  of two-dimensional layered structures through their twist angle}.
\newblock \emph{\bibinfo{journal}{Phys. Rev. B Condens. Matter}}
  \textbf{\bibinfo{volume}{95}}, \bibinfo{pages}{075420}
  (\bibinfo{year}{2017}).

\end{thebibliography}
\bibliographystyle{naturesaa.bst}

\begin{footnotesize}
\vspace{6pt}
\noindent \textbf{Acknowledgment}\\
The authors thank Jicheng Jin, Guixiong Zhong, Stephen Carr, Yuan Cao, Zihao Chen, and Shang Liu for their discussions. The Harvard University team acknowledges support from DARPA under contract URFAO: GR510802. The sample fabrication was performed at Harvard University’s Center for Nanoscale Systems, which is a member of the National Nanotechnology Coordinated Infrastructure Network and is supported by the National Science Foundation under NSF award 1541959. S. F. acknowledges the support of a MURI grant from the U. S. Air Force Office of Scientific Research (Grant No. FA9550-21-1-0312).
\noindent

\vspace{6pt}
\noindent \textbf{Author contributions}\\
Several people contributed to the work described in this paper. H.T. conceived the basic idea for this work. H.T. and R.J., carried out the fabrication. F.D., H.T., and W.X, built the setup and carried out the measurement. F.D., and M.J. carried out the measurement imaging and data processing.  B.L., M.J., X.N., carried out the RCWA simulation. B.L., M.J., carried out the Hamiltonian calculation. E.M., S.F., and H.T. supervised the research and development of the manuscript.  H.T. wrote the first draft of the manuscript and supplementary materials; M.J., F.D., H.T., X.N., and B.L. prepared the first draft of the figures.  H.T., M.J., F.D., X.N. B.L., R.J., E.M., and S.F. worked on the manuscript together. all authors subsequently took part in the revision process, approved the final copy, and provided feedback on the manuscript throughout its development.
\noindent 

\vspace{6pt}
\noindent \textbf{Conflict of interest}\\ The authors declare no competing financial interests.

\noindent 

\vspace{6pt}
\noindent\textbf{Supplementary information}
\noindent is available in the online version of the paper. 
\end{footnotesize}

\end{document}